\documentclass[fleqn,10pt]{wlscirep}

\usepackage[utf8]{inputenc}
\usepackage[T1]{fontenc}
\title{SSDLabeler: Realistic semi-synthetic data generation for multi-label artifact classification in EEG}

\author[1,*]{Taketo Akama}
\author[1,*]{Akima Connelly}
\author[1]{Shun Minamikawa}
\author[1]{Natalia Polouliakh}
\affil[1]{Sony Computer Science Laboratories, Inc., Tokyo, Japan}

\affil[*]{These two authors contributed equally to this work}


\begin{abstract}
EEG recordings are inherently contaminated by artifacts such as ocular, muscular, and environmental noise, which obscure neural activity and complicate preprocessing. Artifact classification offers advantages in stability and transparency, providing a viable alternative to ICA-based methods that enable flexible use alongside human inspections and across various applications. However, artifact classification is limited by its training data as it requires extensive manual labeling, which cannot fully cover the diversity of real-world EEG. Semi-synthetic data (SSD) methods have been proposed to address this limitation, but prior approaches typically injected single artifact types using ICA components or required separately recorded artifact signals, reducing both the realism of the generated data and the applicability of the method. To overcome these issues, we introduce SSDLabeler, a framework that generates realistic, annotated SSDs by decomposing real EEG with ICA, epoch-level artifact verification using RMS and PSD criteria, and reinjecting multiple artifact types into clean data. When applied to train a multi-label artifact classifier, it improved accuracy on raw EEG across diverse conditions compared to prior SSD and raw EEG training, establishing a scalable foundation for artifact handling that captures the co-occurrence and complexity of real EEG.

\end{abstract}
\begin{document}

\flushbottom
\maketitle

\thispagestyle{empty}

\section*{Introduction}

Electroencephalography (EEG) is a widely used, non-invasive neuroimaging technique that measures brain activity from the scalp. EEG's high temporal resolution, portability, and relatively low cost have made it indispensable in neuroscience research, clinical diagnostics, machine learning applications, and brain–computer interface (BCI) development \cite{lotte_review_2018,craik_deep_2019,roy_deep_2019}. EEG signals reflect small electrical potentials generated by synchronized neuronal firing, yet the sensitivity required to detect these weak signals also makes EEG highly susceptible to contamination from non-neural sources, including ocular and muscular activity, head movements, electrode fluctuations, and environmental interference. These artifacts frequently overlap with genuine brain signals in both time and frequency, degrading the signal-to-noise ratio (SNR) and reducing the reliability of downstream analyses such as neural decoding and event-related feature extraction. Despite the diversity of EEG applications, including event-related investigations in neuroscience \cite{zhao_hand_2019,liu_detecting_2017}, clinical diagnostics, and machine-learning-based classification \cite{gu_eeg-based_2021,rashid_current_2020,cai_feature-level_2020,nicolas-alonso_brain_2012,herrmann_human_2005}, all users share the need to preserve meaningful neural activity while controlling for contaminating artifacts. Achieving this balance requires preprocessing methods that enhance interpretability without distorting or removing genuine neural events.

Current artifact handling strategies, however, face several limitations. Manual inspection remains common but is subjective, labor-intensive, and reliant on the experience of the reviewer \cite{uriguen_eeg_2015}. Independent Component Analysis (ICA) offers a semi-automated alternative, and ICLabel is widely adopted for classifying ICA components into brain, eye, muscle, heart, line, channel, or other categories \cite{naeem_seperability_2006,kanoga_assessing_2016,pion-tonachini_iclabel_2019,makeig_independent_1995,hyvarinen_independent_2000}. Yet ICA-based artifact removal eliminates entire components, which risks discarding neural signals that co-occur with artifacts. Furthermore, ICA performance is sensitive to data quality, channel count, and subject variability, limiting its reliability and necessitating expert verification \cite{delorme_eeg_2023}. More recent deep learning–based denoising models can produce end-to-end “cleaned” EEG but operate as black boxes \cite{zhang_eegdenoisenet_2021,chuang_ic-u-net_2022}. They lack transparent artifact detection and may inadvertently remove meaningful neural features, including event-related potentials (ERPs). Their performance is further constrained by the limited availability and variability of artifact-labeled training data. Earlier artifact classification approaches relying on principle component analysis (PCA), ICA, or wavelet features were similarly limited by the scarcity of labeled data and generally underperformed compared with contemporary approaches \cite{chaddad_electroencephalography_2023,jiang_removal_2019,islam_methods_2016}. Nevertheless, artifact classification retains a key advantage: transparency and user control, enabling researchers to decide how to handle identified contamination rather than relying on fully automated removal.

Most existing artifact detection approaches operate under a binary “clean versus artifact” assumption, which overlooks the fact that different artifact types frequently co-occur and can have distinct effects on downstream analysis. To address this, we propose a multi-label artifact classification framework that identifies multiple artifact types (eye, muscle, heart, line, channel, and other noise) within a single time-locked EEG segment (epoch). This provides transparent, per-epoch contamination information and offers users flexibility in how they handle flagged data. Supporting multi-label classification requires a diverse and realistic set of labeled examples, however, artifact-labeled datasets are scarce and inherently limited in their coverage of real-world variability.

To overcome this constraint, we introduce a novel semi-synthetic data (SSD) generation method that produces realistic multi-label contaminations reflective of those observed in raw EEG. Prior SSD approaches have primarily injected single artifact types into clean EEG, producing isolated contaminations that fail to capture the co-occurrence structure \cite{zhang_eegdenoisenet_2021,chuang_ic-u-net_2022}. In contrast, our framework reconstructs artifact sources from ICA-derived components and uses root-mean-square (RMS) and power spectral density (PSD) criteria to verify the presence of artifacts at the epoch level before reinjection. By simultaneously mixing multiple artifact components, our SSD preserves both spatiotemporal artifact signatures and underlying brain activity, generating realistic multi-label contamination suitable for training robust classifiers.

Beyond improving classification, the proposed approach enables several practical applications. It supports objective, automated reporting for trial or epoch rejection, reducing subjectivity and improving reproducibility across studies. It can be applied prior to ICA to exclude highly contaminated segments, improving ICA stability and producing more reliable component labeling \cite{klug_optimizing_2024,callan_shredding_2024}. It also facilitates selective cleaning in averaging-based analyses such as ERP or PSD, preventing contaminated epochs from distorting neural feature estimates. Motivated by these applications, the present study evaluates how multi-label SSD generation, combined with multi-label artifact classification, can support a scalable, transparent, and researcher-centered framework for EEG preprocessing—one that preserves neural information while providing actionable, interpretable artifact detection.

\section*{Results}

\subsection*{Multi-label artifact classification performance}
We evaluated multi-label artifact classification performance using data from five subjects that were held out from training and validation. Three models were trained separately using (i) raw EEG, (ii) our proposed SSD-generated dataset, and (iii) the SSD baseline from prior work. All datasets were annotated using our SSDLabeler procedure. The resulting models were then tested on expert-annotated data from both the motor execution (ME) and noise sessions to provide an unbiased assessment of generalization performance across artifact types.

\subsubsection*{Motor execution session}
On the ME test dataset, our proposed SSD method achieved the highest overall accuracy ($0.839$), surpassing both raw EEG ($0.772$) and the SSD baseline from previous work ($0.788$). Category-wise performance is summarized in Table~\ref{tab:table_ME_acc}. Our SSD achieved the highest accuracies for the \textit{Clean} ($0.850$), \textit{Eye} ($0.783$), \textit{Heart} ($1.0$), \textit{Line} ($1.0$), \textit{Channel} ($0.833$), and \textit{Other} ($0.803$) categories. In contrast, the previous work's SSD achieved the best performance for the \textit{Muscle} category ($0.785$).

\begin{figure}[ht]
\centering
\includegraphics[width=0.9\linewidth]{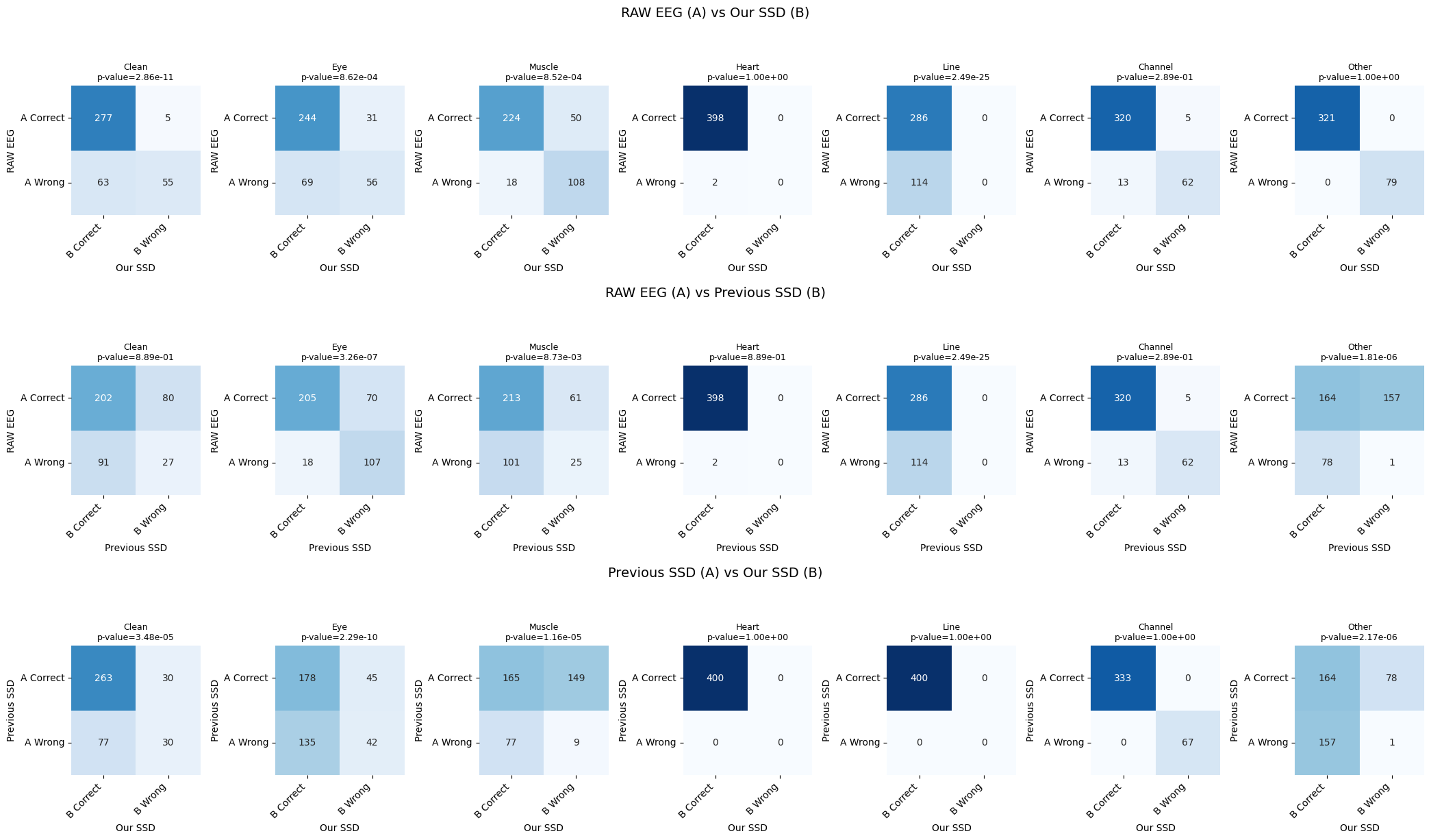}
\caption{McNemar's test results for per-label classification performance across the three training datasets on the motor execution session test data. The top row (a) compares Raw EEG with Our SSD, the middle row (b) compares Raw EEG with the Previous Work's SSD, and the bottom row (c) compares the Previous Work's SSD with Our SSD. In all panels, columns correspond to artifact categories in the following order: Clean, Eye, Muscle, Heart, Line, Channel, and Other.}

\label{fig:fig_McNemar_ME}
\end{figure}

To examine whether these performance differences reflected statistically meaningful changes in error distributions, we conducted McNemar’s tests for each artifact label, comparing (i) our SSD with raw EEG, (ii) our SSD with the previous work’s SSD, and (iii) raw EEG with the previous work’s SSD. Holm-corrected p-values indicated significant improvements of our SSD over raw EEG for the \textit{Clean}, \textit{Eye}, and \textit{Line} categories ($p < 0.05$) seen in Figure~\ref{fig:fig_McNemar_ME} (a). A significant difference was also observed for the \textit{Muscle} category, where raw EEG outperformed our SSD.

When comparing raw EEG and the previous work’s SSD, significant differences were revealed in the \textit{Eye}, \textit{Muscle}, \textit{Line}, and \textit{Other} categories ($p < 0.05$). Raw EEG outperformed the previous work’s SSD in the \textit{Eye} and \textit{Other} categories, whereas the previous work’s SSD outperformed raw EEG in the \textit{Muscle} and \textit{Line} categories. The complete set of McNemar’s test results is visualized in Fig.~\ref{fig:fig_McNemar_ME} (b).

Finally, comparisons between our SSD with the previous work's SSD, significant improvements were observed for the \textit{Clean}, \textit{Eye}, and \textit{Other} categories ($p < 0.05$) seen in Figure~\ref{fig:fig_McNemar_ME} (c). However, the previous work's SSD significantly outperformed our method for the \textit{Muscle} category.
\begin{table}[ht]
\centering

\caption{Classification accuracy by artifact category in the ME session (human-labeled test set).}
\label{tab:table_ME_acc}
\begin{tabular}{lccclclcc}
\hline
Method & Clean & Eye & Muscle  & Heart & Line& Channel& Other & Average \\
\hline
Raw EEG& 0.813& 0.688& 0.685 &0.995& 0.715&0.813& 0.803& 0.772\\
Previous Work's SSD\cite{chuang_ic-u-net_2022}            & 0.733& 0.558& 0.785&1.0& 1.0&0.833& 0.603& 0.788\\
Our SSD& 0.850& 0.783& 0.605&1.0& 1.0&0.833& 0.803& 0.839\\
\hline
 SSDLabeler & 0.860& 0.773& 0.600& 1.0& 1.0& 0.833& 0.803&0.838\\
\end{tabular}
\end{table}

\subsubsection*{Noise session}
On the noise session containing instructed artifacts, our proposed SSD method achieved the highest overall accuracy ($0.812$), outperforming both raw EEG ($0.618$) and the previous work’s SSD ($0.756$). Category-wise results are shown in Table~\ref{tab:table_noise_acc}. Our SSD achieved the highest accuracies for the \textit{Clean} ($1.0$), \textit{Eye} ($0.871$), \textit{Heart} ($1.0$), \textit{Line} ($1.0$), \textit{Channel} ($0.965$), and \textit{Other} ($0.482$) categories. As in the ME session, the previous work’s SSD achieved the best performance for the \textit{Muscle} category ($0.624$).

\begin{figure}[ht]
\centering
\includegraphics[width=0.9\linewidth]{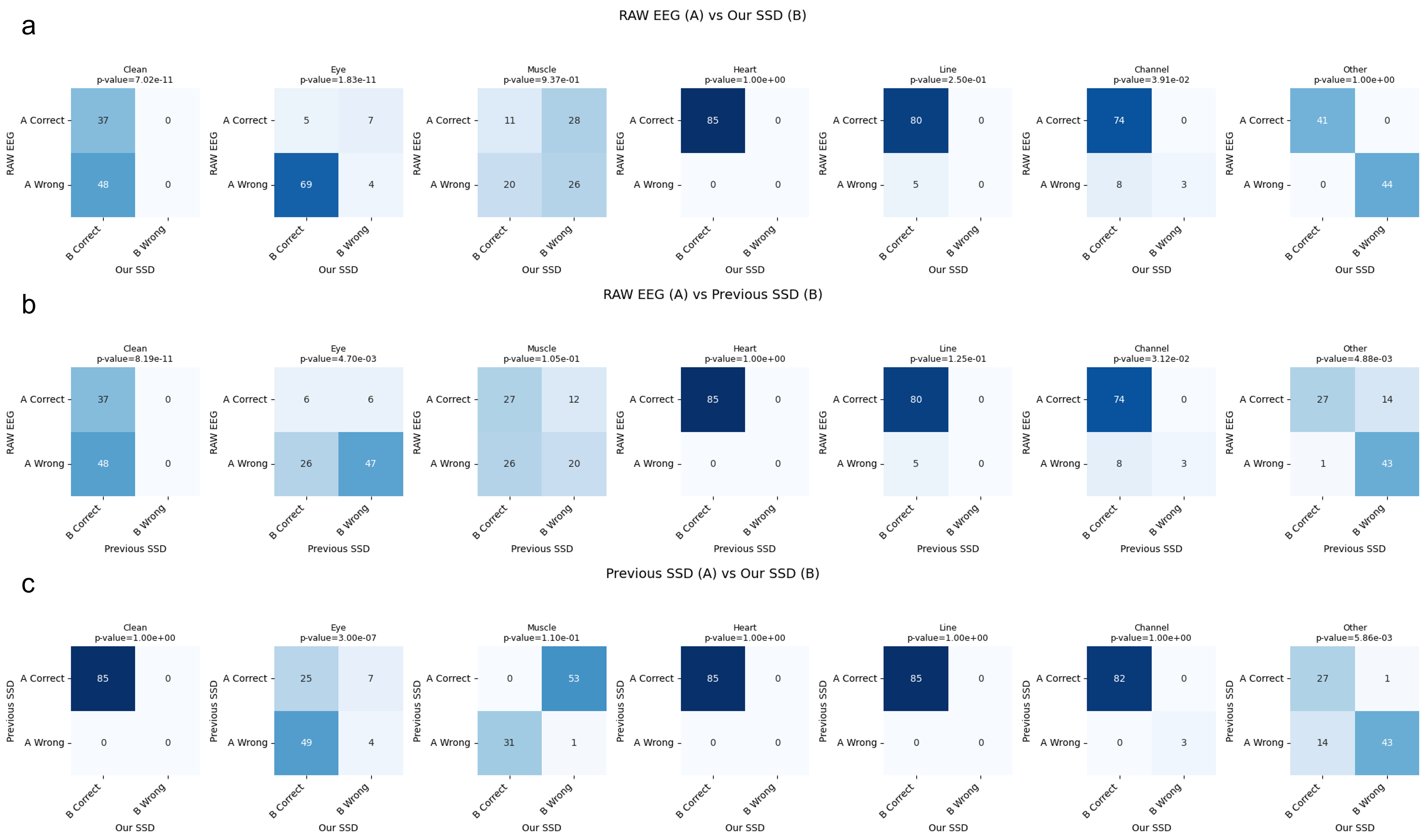}
\caption{McNemar's test results for per-label classification performance across the three training datasets on the noise session test data. The top row (a) compares Raw EEG with Our SSD, the middle row (b) compares Raw EEG with the Previous Work's SSD, and the bottom row (c) compares the Previous Work's SSD with Our SSD. In all panels, columns correspond to artifact categories in the following order: Clean, Eye, Muscle, Heart, Line, Channel, and Other.}

\label{fig:fig_McNemar_Noise}
\end{figure}

To evaluate whether these differences reflected significant changes in error patterns, McNemar’s tests with Holm correction were applied to each artifact label. When comparing our SSD with raw EEG, significant improvements were observed for the \textit{Clean}, \textit{Eye}, and \textit{Channel} categories ($p < 0.05$), indicating that our SSD outperformed raw EEG in these specific artifact types as seen in Figure~\ref{fig:fig_McNemar_Noise} (a).

Comparisons between raw EEG and the previous work’s SSD, significant differences were observed for the \textit{Clean}, \textit{Eye}, \textit{Channel}, and \textit{Other} categories ($p < 0.05$) seen in Figure~\ref{fig:fig_McNemar_Noise} (b). The previous work’s SSD outperformed raw EEG for the \textit{Clean}, \textit{Eye}, and \textit{Channel} categories, whereas raw EEG outperformed the previous work’s SSD for the \textit{Other} category.

Finally, when comparing our SSD and the previous work’s SSD revealed significant differences for the \textit{Eye} and \textit{Other} categories ($p < 0.05$), with our SSD showing superior performance in both cases seen in Figure~\ref{fig:fig_McNemar_Noise} (c).

\begin{table}[ht]
\centering
\caption{Classification accuracy by artifact category in the Noise session (human-labeled test set).}
\label{tab:table_noise_acc}
\begin{tabular}{lccclllcc}
\hline
Method & Clean & Eye & Muscle   &Heart& Line&Channel& Other & Average \\
\hline
Raw EEG (SSD annotated) & 0.435& 0.141& 0.459  &1.0& 0.941&0.871& 0.482& 0.618\\
Previous Work's SSD\cite{chuang_ic-u-net_2022}            & 1.0& 0.377& 0.624&1.0& 1.0&0.965& 0.329& 0.756\\
Our SSD& 1.0& 0.871& 0.365&1.0& 1.0&0.965& 0.482& 0.812\\

\hline
\end{tabular}
\end{table}

\section*{Discussion}
This study introduced a multi-label artifact classification framework supported by a novel ICA-based SSD generation method designed to overcome the persistent shortage of artifact-labeled EEG data. A major challenge in EEG preprocessing is that artifacts rarely occur in isolation; instead, they frequently co-occur, overlap in time, and exhibit nonstationary dynamics that complicate both manual inspection and automated removal. Existing ICA-based SSD approaches address these challenges only partially, as they typically inject a single artifact type into clean EEG \cite{chuang_ic-u-net_2022}, producing simplified examples that do not adequately represent the complex contamination patterns present in real EEG. By contrast, our proposed SSD method reintroduces multiple ICA-isolated artifact components simultaneously, preserving their inherent spatiotemporal structure and producing a richer, more ecologically valid representation of artifact mixtures.

The results of this study highlight two key advantages of our approach: (i) improved realism in synthetic contamination and (ii) enhanced training effectiveness for multi-label artifact classification. When trained on our SSD, the multi-label classifier demonstrated the highest accuracy on the human-annotated ME session, outperforming both raw EEG and the previous SSD baseline across most artifact categories. These findings suggest that the contamination patterns generated by our method more closely resemble real-world EEG, enabling the classifier to generalize more reliably to naturally occurring artifacts. Importantly, improvements were not restricted to any single artifact type; rather, the SSD-supported classifier showed consistently balanced performance, indicating that the inclusion of multiple artifact components during training provides meaningful diversity needed for robust multi-label prediction.

A second validation was performed using an instructed-noise session, in which participants intentionally generated artifact-specific behaviors. This dataset, characterized by clearly defined and high-intensity contaminations, serves as a strong benchmark for evaluating artifact classifiers under controlled yet realistic conditions. On this dataset, our SSD-trained model again achieved the best overall accuracy, demonstrating strong advantages in detecting clean, eye-related, line-related, cardiac, and channel noise artifacts. The only notable limitation was reduced performance in muscle artifact detection, a pattern observed in both sessions. This suggests that ICA extraction of muscle components may be less consistent or that muscle artifacts inherently possess more inter-trial variability, motivating future work to incorporate complementary EMG-informed features, time-frequency representations, or data-driven muscle priors to better capture their diverse morphology.

The broader implication of this work is that SSD, when designed to reflect the co-occurrence structure of real EEG, can substantially enhance the performance of artifact-aware machine learning pipelines. Our method demonstrates that synthetic augmentation is not merely a workaround for limited labeled data, but can be an integral part of a principled strategy for training generalizable artifact classifiers. This is particularly important for large-scale or real-time EEG applications, where automated and reliable artifact detection is essential but manually labeled data is difficult or impossible to obtain at scale. Moreover, because our SSD generation is fully automated and does not require manual inspection, it offers a scalable solution for producing high-quality training data across diverse experimental paradigms, EEG systems, and subject populations.

Despite the promising results, several limitations remain. Muscle artifact performance indicates that certain artifact classes may require additional modeling beyond ICA-derived components. Furthermore, although our method preserves the spatial maps and temporal waveforms of ICA components, future research could explore combining SSD with more physiologically grounded generative approaches or hybrid models that integrate time-frequency features or deep generative neural networks. Finally, while this study focused on multi-label classification as a downstream task, future work may evaluate how SSD-augmented preprocessing influences source localization or neural decoding tasks relevant to BCI and cognitive neuroscience.

Overall, the findings demonstrate that our multi-artifact SSD generation method substantially improves the realism and diversity of synthetic contaminated EEG, translating into enhanced performance of a multi-label artifact classifier under both naturalistic and experimentally controlled conditions. These results establish the proposed approach as a valuable and scalable contribution toward fully automated, artifact-aware EEG preprocessing pipelines.

\section*{Methods}
\subsection*{Dataset and Preprocessing}
We used the publicly available EEG dataset introduced by Cho et al.\, which contains recordings from motor execution (ME), motor imagery (MI), rest, and an instructed-noise session \cite{cho_eeg_2017}. EEG was acquired from 52 participants using a 64-channel BioSemi ActiveTwo system at a sampling rate of 512 Hz. For the present study, only the ME sessions were used to construct the training and evaluation sets for artifact classification. Data from 42 subjects were used for training, 5 for validation, and 5 for an independent test set. The validation and test data consisted only of raw and cleaned EEG.

Each ME session consisted of 20 trials per hand. At the start of each trial, subjects viewed a black screen followed by a fixation cross presented for two seconds. This was followed by a randomly ordered instruction cue (“left hand” or “right hand”) presented for three seconds, during which participants executed the cued motor movement. The three-second movement period was extracted for all analyses in this study.

In addition to the task-related sessions, the Cho et al.\ dataset also includes a dedicated noise session designed to capture instructed artifact behaviors. Five types of non-task-related activities were recorded: eye blinking, vertical eye movements (up/down), horizontal eye movements (left/right), head movements, and jaw clenching. For each artifact type, participants performed the instructed movement twice for 5 seconds. These noise recordings contain clearly observable, high-amplitude artifacts, providing a complementary dataset for evaluating artifact classification under controlled conditions and benchmarking the behavior of models trained on semi-synthetic contamination.

Preprocessing followed a modified version of the pipeline described by Zhang et al.\ \cite{zhang_eegdenoisenet_2021}. Raw EEG was band-pass filtered between 1 and 50 Hz using a zero-phase finite impulse response (FIR) filter of order 6600 implemented with the EEGLAB \texttt{filter\_data} function \cite{delorme_eeglab_2004}. Power-line interference at 60 Hz was suppressed using the \texttt{filter\_notch} function. After filtering, continuous data were detrended to remove slow drifts and segmented into three-second epochs aligned to cue onset. Finally, the data were downsampled from 512 Hz to 125 Hz, following common EEG preprocessing practice and ensuring compatibility with downstream model architectures.

\subsection*{ICA and Artifact IC Selection}
Independent Component Analysis (ICA) was applied to the preprocessed EEG to separate the multichannel signals into temporally independent and spatially fixed components. ICA is widely used for isolating neural sources from ocular, muscular, cardiac, and environmental artifacts by exploiting statistical independence \cite{naeem_seperability_2006,kanoga_assessing_2016,makeig_independent_1995}. Prior to decomposition, all epochs were concatenated across trials to form a continuous matrix suitable for ICA. Specifically, the epoched data of size $(C \times T \times N)$, with $C=64$ channels, $T=375$ samples, and $N$ trials, were reshaped into a matrix of size $C \times (T\cdot N)$, ensuring that ICA operated on the full dataset rather than on individual segments. Principal component analysis (PCA) whitening was then performed. FastICA was subsequently applied to estimate the independent components \cite{hyvarinen_independent_2000}.

\begin{figure}[ht]
\centering
\includegraphics[width=\textwidth]{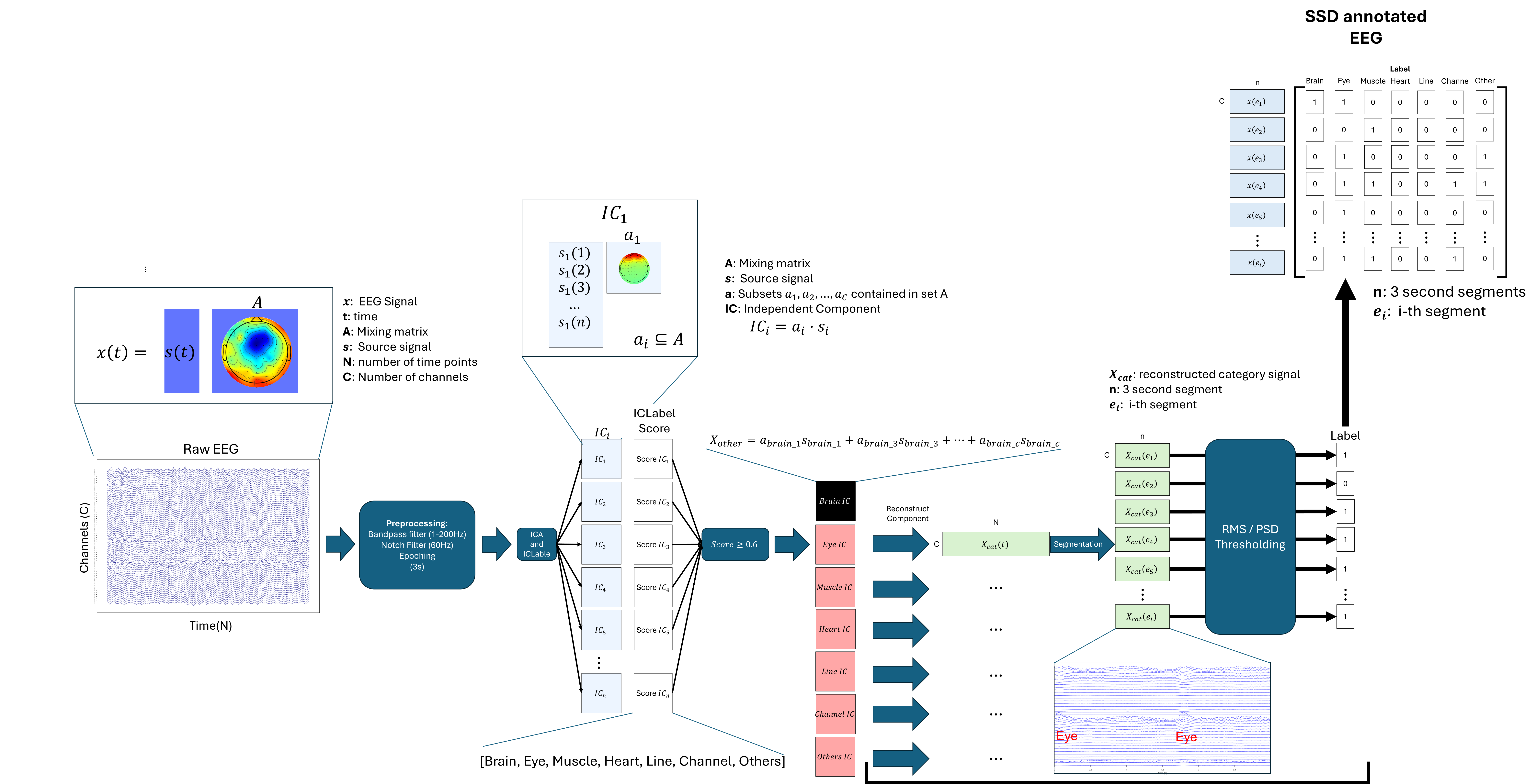}
\caption{Preprocessing and annotation pipeline for semi-synthetic EEG generation. Raw EEG was first preprocessed with bandpass (1–50 Hz) and notch filtering (60 Hz), then epoched into three-second segments. ICA and ICLabel were then applied to decompose the signals and assign artifact probabilities to each IC. Artifact ICs with ICLabel scores $\geq 0.6$ (e.g., eye, muscle, heart, line, channel, or other) were reconstructed and segmented into time windows. RMS and PSD thresholding were subsequently applied to ensure the presence of true contamination within each segment. This pipeline enables reliable annotation of artifact-contaminated EEG segments, forming the basis for generating labeled, realistic, multi-label SSD.}

\label{fig:fig5}
\end{figure}

Artifact-related components were identified using ICLabel, an automated classifier trained to assign probabilistic labels to ICA components \cite{pion-tonachini_iclabel_2019}. ICLabel outputs probabilities across seven categories: brain, eye, muscle, heart, line noise, channel noise, and other. Components with an artifact probability $\geq 0.6$ for any non-brain category were retained as artifact ICs. Components with a dominant brain label were kept for reconstructing clean EEG, while components that did not strongly match either category were discarded to avoid ambiguous source mixing. This procedure produced a structured library of artifact ICs—ocular, muscular, cardiac, line, channel noise, and other noise—that served as the basis for semi-synthetic EEG generation, while ensuring that neural activity reconstructed from the brain-labeled ICs remained as clean as possible. Figure~\ref{fig:fig5} summarizes the preprocessing and IC annotation pipeline.

\subsection*{SSD Labeler for Artifact Annotation}
Although ICLabel provides probabilistic classifications of independent components, some IC-derived artifact epochs may still be ambiguous or may not contain meaningful artifact activity. To ensure that only segments exhibiting clear, physiologically plausible contamination were used for SSD generation, we developed an additional verification tool, \textit{SSDLabeler}, that integrates ICLabel outputs with signal-based thresholding criteria. SSDLabeler performs a secondary screening based on RMS and PSD features to refine the detection of artifact-containing epochs.

RMS thresholding is applied to capture transient, high-amplitude non-neural events—including ocular activity, cardiac activity, line noise, channel noise, and other abrupt contaminations by evaluating signal energy relative to percentile-based thresholds computed within each IC. In contrast, PSD thresholding is used to identify muscle-related artifacts, which typically exhibit elevated high-frequency spectral power. By detecting increases in power within a muscle-specific frequency range, PSD verification filters out IC segments with weak or absent muscle contamination.

By combining data-driven ICLabel probabilities with adaptive RMS- and PSD-based verification, SSDLabeler ensures that only reliably contaminated IC segments are selected for reinjection into clean EEG. This additional screening step reduces false positives from ICLabel alone and yields artifact annotations that more accurately reflect the temporal and spectral characteristics of real EEG contamination, thereby supporting the generation of a higher-quality and more realistic semi-synthetic EEG dataset.

\subsubsection*{RMS thresholding}
The RMS value provides a measure of the overall signal energy within an epoch. In Eq.~(\ref{eq:rms}), $x$ denotes the referenced signal of length $N$ used to compute the RMS:

\begin{equation}
\mathrm{RMS}(x) = \sqrt{\frac{1}{N} \sum_{i=1}^{N} x_i^2}.
\label{eq:rms}
\end{equation}
To quantify artifact strength, a standardized effect score was computed (Eq.~\ref{eq:standardize}) that compares the mean RMS of artifact-labeled epochs, $\mu(RMS_{\text{artifact}})$, with that of clean EEG, $\mu(RMS_{\text{clean}})$, normalized by the standard deviation of the clean RMS distribution, $\sigma(RMS_{\text{clean}})$:

\begin{equation}
RMS_z = \frac{\mu(RMS_{\text{artifact}}) - \mu(RMS_{\text{clean}})}{\sigma(RMS_{\text{clean}})}.
\label{eq:standardize}
\end{equation}

Thresholds for each artifact category were defined by taking the $N^\text{th}$ percentile of all $RMS_z$ scores (Eq.~\ref{eq:RMS_Threshold}), enabling the selection of the strongest artifact instances during RMS screening. The percentile thresholds used were: \textit{Eye} (0.1), \textit{Heart} (0.99), \textit{Line} (0.99), \textit{Channel} (0.99), and \textit{Other} (0.99).

\begin{equation}
Threshold = Percentile(RMS_z, x).
\label{eq:RMS_Threshold}
\end{equation}

\begin{figure}[ht]
\centering
\includegraphics[width=0.6\linewidth]{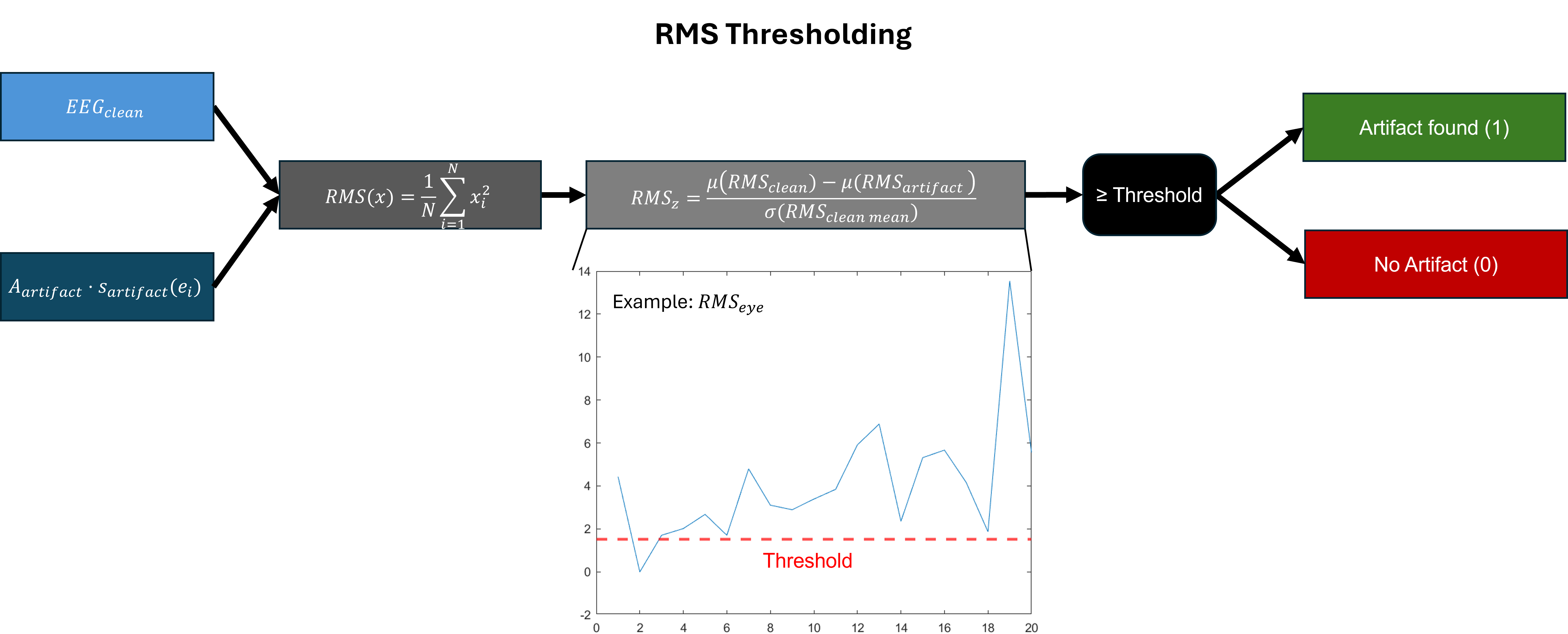}
\caption{RMS thresholding procedure used for artifact verification. For all artifact types except muscle, IC-derived signals were segmented into epochs and the RMS was computed for each segment. Epochs with RMS values exceeding the predefined threshold were labeled as artifact-present (1), whereas those below the threshold were labeled as clean (0). The example plot illustrates $RMS_{\text{eye}}$, where segments above the red dashed line are identified as eye-related artifacts.}

\label{fig:fig_rms}
\end{figure}

Epochs whose RMS scores fell below their respective category threshold were excluded from SSD generation, ensuring that only IC segments containing clear artifact activity were reinjected into clean EEG (Figure~\ref{fig:fig_rms}). RMS thresholding was particularly effective for identifying transient, high-amplitude non-neural events; consequently, all artifact types except muscle were screened using this method.

\subsubsection*{PSD thresholding}
\begin{figure}[ht]
\centering
\includegraphics[width=0.6\linewidth]{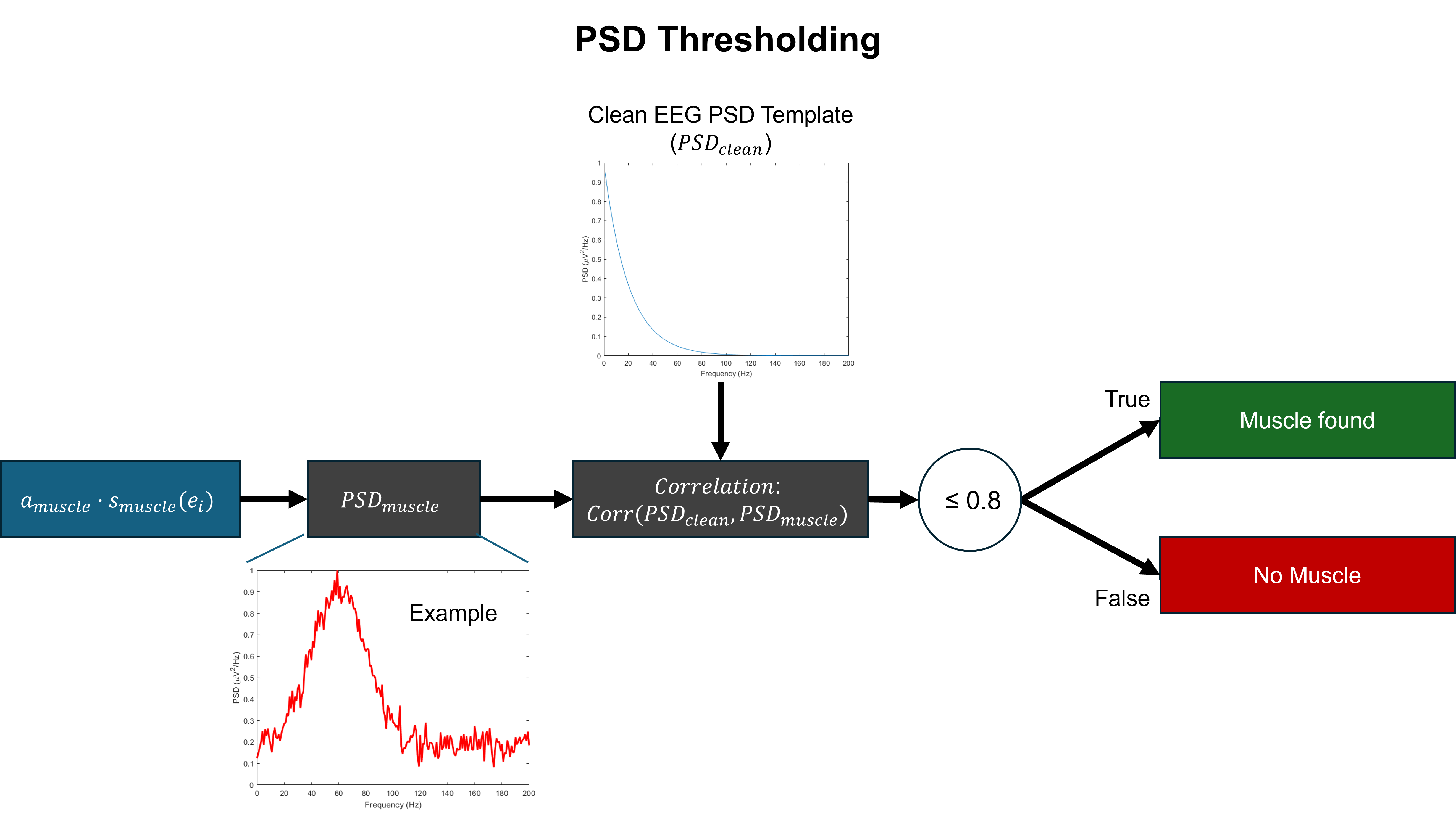}
\caption{PSD thresholding procedure for muscle artifact detection. The PSD of muscle component segments and correlation values were computed. 
Segments with correlation $\leq 0.8$ relative to the clean PSD template were classified as containing muscle artifacts (1), while those above the threshold were labeled as clean (0). A simulated muscle spectrum ($PSD_{\text{muscle}}$) was used as an example to show elevated high-frequency activity characteristic of muscle contamination.}

\label{fig:fig_psd}
\end{figure}

PSD thresholding was used to label muscle artifacts within the epochs, as they are typically represented by large bursts in power in the high-frequency range. Following Zhang et al., we constructed a clean EEG PSD template and compared it with the PSDs of muscle ICs using correlation analysis\cite{zhang_eegdenoisenet_2021}. Unlike Zhang et al., who rejected low-correlation segments, we retained these segments as verified muscle artifacts. Epochs with correlation $\leq 0.8$ relative to the clean template were labeled as muscle artifact (1), while those above were discarded as seen in Figure \ref{fig:fig_psd}.

\subsubsection*{SSDLabeler Hyperparameter Optimization}

To determine the optimal parameters for SSDLabel, we performed a structured hyperparameter search that jointly evaluated the ICLabel probability threshold and the class-specific RMS and PSD verification thresholds. The objective was to identify a set of thresholds that maximized artifact detection accuracy while maintaining consistent performance across all recordings. As ICLabel assigns each independent component (IC) a probability distribution over seven classes, the confidence required for accepting an IC as an artifact can influence which components contribute to the reconstructed artifact signals \cite{pion-tonachini_iclabel_2019}. We therefore evaluated a set of ICLabel probability thresholds,
\begin{equation}
p \in \{0.50,\,0.60,\,0.70,\,0.80,\,0.90\},
\label{eq:ICLabel_hyperparameter}
\end{equation}
such that an IC was assigned to its ICLabel-predicted artifact class only if its winning probability exceeded \(p\). ICs whose winning probability was below the threshold were excluded. For each value of \(p\), this procedure yielded a different IC-to-class mapping and a corresponding reconstructed artifact signal for each artifact category.

After the IC assignment, the per-epoch RMS-based standardized scores were computed for non-muscle artifacts and the PSD-correlation values for muscle artifacts (see previous subsections). To avoid evaluating the full combinatorial space of all parameters, we performed a one-parameter-at-a-time sweep for each artifact class while reusing the same cached feature matrices.

For the five artifact categories screened with RMS (\emph{eye}, \emph{heart}, \emph{line noise}, \emph{channel noise}, \emph{other}), we evaluated using percentile scores to define the threshold,
\begin{equation}
\theta \in \{10,\,20,\,30,\,40,\,50,\,60,\,70,\,75,\,80,\,85,\,90,\,95,\,99\},
\label{eq:rms_hyperparameter}
\end{equation}
and selected the value that maximized the per-class accuracy when comparing the predicted labels against the ground-truth annotations for that category. For \emph{muscle} artifacts, which were evaluated using PSD correlation, we swept the correlation thresholds
\begin{equation}
\tau \in \{0.60,\,0.70,\,0.80,\,0.85,\,0.90\},
\label{eq:psd_hyperparameter}
\end{equation}
and selected the value that produced the highest muscle artifact detection accuracy.
For each ICLabel probability threshold \(p\), the sweeps above produced a set of optimal parameters,
\begin{equation}
\{\theta^\ast_{\mathrm{eye}},\,
\theta^\ast_{\mathrm{heart}},\,
\theta^\ast_{\mathrm{line}},\,
\theta^\ast_{\mathrm{channel}},\,
\theta^\ast_{\mathrm{other}},\,
\tau^\ast_{\mathrm{muscle}}\}.
\label{eq:best_hyperparameter}
\end{equation}
To determine the final configuration, we computed the macro-accuracy (Eq.~\ref{eq:mean_macro_accuracy}), defined as the mean classification accuracy across all seven artifact classes, for each tested value of \(p\). Ground-truth labels used for this evaluation were human-annotated by an EEG expert. The ICLabel probability threshold that produced the highest macro-accuracy across recordings was selected as the final operating point for SSDLabeler, together with its corresponding set of class-specific hyperparameters. This data-driven optimization procedure ensured a principled selection of parameters that balanced artifact sensitivity and specificity while maintaining robustness across subjects and recording sessions.

\begin{equation}
\text{Mean Macro Accuracy} =
\frac{1}{N_{\text{files}} \times N_{\text{cat}}}
\sum_{i=1}^{N_{\text{files}}}
\sum_{c=1}^{N_{\text{cat}}}
Acc_{i,c},
\label{eq:mean_macro_accuracy}
\end{equation}

\subsection*{Semi-Synthetic Data Generation (SSDLabeler)}
  
We developed a SSD generation framework, termed \textit{SSDLabeler}, to produce realistic multi-label artifact–contaminated EEG. Similar to the previous single-artifact SSD approach, ICA and ICLabel were used to classify independent components into artifact categories based on their probabilistic labels. To maintain consistent conditions across methods, the ICLabel probability threshold was set to $\geq 0.6$ for both our SSD and the previous work’s SSD implementation. The previous approach originally used a 0.8 threshold, which we adjusted to match our optimized hyperparameter settings and ensure a fair comparison.

The key difference between prior SSD methods and our proposed framework lies in the handling of artifact components. Previous SSD approaches inject only a single artifact category into clean EEG, producing isolated contaminations that do not reflect the co-occurrence structure of real-world artifacts. In contrast, SSDLabeler incorporates an additional verification step: only artifact IC epochs that pass the RMS- or PSD-based thresholding criteria are selected for reinjection. This ensures that each IC segment contains verifiable ocular, muscular, cardiac, line, channel, or other artifact activity.

The verified artifact ICs are then mixed back into clean EEG to create multi-artifact contaminated signals, while preserving both the underlying neural activity and the spatiotemporal characteristics of the artifact components. This process generates realistic multi-label contamination that more closely resembles the complexity and overlapping structure of raw EEG recordings.

\subsection*{Model}
To evaluate the utility of the SSD, we trained a deep neural network for multi-label artifact classification. The classifier architecture was adapted from the EEG encoder branch of the PredANN model, originally developed for music classification using EEG and audio through contrastive learning \cite{akama_predicting_2025}. For the present study, the original classification module was used and modified with a multi-label prediction head, enabling the model to assign multiple artifact labels to each EEG segment. In this streamlined configuration, the model focuses exclusively on identifying artifact-contaminated EEG based on its spatiotemporal structure.

\subsubsection*{Model Architecture and Losses}
The adapted model consists of a two-dimensional convolutional neural network (2D CNN) designed to extract hierarchical representations from multichannel EEG. Inputs were formatted as matrices of size $C \times T$, where $C = 64$ EEG channels and $T = 375$ time samples correspond to three-second epochs sampled at 125~Hz. The encoder is composed of multiple convolutional blocks with kernel sizes selected to capture spatial dependencies across channels as well as temporal dynamics within each trial. Each block contains convolutional layers followed by batch normalization, nonlinear activation functions, and dropout for regularization. The resulting feature maps are flattened and passed to a fully connected classification head with sigmoid activation, allowing independent probability estimates for each artifact category (eye, muscle, heart, line, channel, and clean).

The model was trained using the binary cross-entropy loss function, consistent with multi-label classification. Labels for each input segment were derived from the SSDLabeler–based annotation framework, ensuring that the training set included realistic examples of artifact contamination. Optimization was performed using the Adam optimizer with an initial learning rate of $1\times10^{-3}$ and a batch size of 48. Training proceeded for 5000 epochs to ensure full convergence and stable multi-label predictions.

\section*{Conclusion}
This study introduced a novel SSD generation framework to address a long-standing limitation in EEG research: the lack of high-quality, artifact-labeled data that captures the complexity, diversity, and co-occurrence structure of real-world contamination. By leveraging ICA-derived artifact components together with RMS- and PSD-based thresholding, the proposed method ensures that only physiologically plausible and temporally valid sources are incorporated into the SSD. Crucially, unlike prior approaches that inject a single artifact type into clean EEG, our method reintroduces multiple independent artifact components concurrently, preserving their spatial projections, temporal morphology, and natural overlap. This multi-artifact design produces a realistic and ecologically valid representation of contamination, allowing the synthetic data to more closely reflect the realistic properties of raw EEG.

The effectiveness of this SSD framework was demonstrated through a multi-label artifact classification task evaluated on both real-world ME recordings and a controlled instructed-noise session with intentionally produced artifacts. Across both settings, the classifier trained on our SSD achieved superior overall accuracy compared with models trained on raw EEG or on the previous SSD baseline. These improvements were observed across nearly all artifact categories, indicating that the realistic variability and multi-artifact interactions captured by our SSD translate directly into stronger generalization performance. Although muscle artifacts remain a challenge, likely due to the heterogeneous nature of EMG contamination and the variability of ICA-derived muscle components, this limitation highlights a promising direction for extending the framework with EMG-informed representations, improved ICA extraction strategies, or hybrid generative models.

Beyond achieving high classification accuracy, the value of the proposed SSD framework extends to broader methodological implications for EEG preprocessing. Realistic multi-artifact synthetic data enables automated and reproducible artifact identification, reducing dependence on subjective manual inspection and enhancing consistency across studies. Improved artifact classification further benefits preprocessing pipelines by supporting pre-ICA quality screening, preserving ICA stability, and facilitating selective cleaning in analyses such as ERP or PSD. These capabilities reflect a shift toward principled, scalable, and automated preprocessing strategies suited for large datasets, diverse populations, and real-time applications where manual annotation is impractical. More broadly, this work demonstrates that SSD, when grounded in physiologically meaningful transformations, can strengthen artifact-aware machine learning models beyond what is possible through raw data alone. By accurately modeling the mixed and overlapping structure of EEG contamination, the proposed SSD provides a foundation for training classifiers that generalize across subjects, experimental paradigms, and recording systems. As EEG datasets continue to expand in scale and complexity, such automated augmentation strategies will play an increasingly important role in ensuring robust and reproducible downstream analyses. Looking ahead, extending the current architecture with a temporal segmentation head capable of generating continuous, pointwise predictions represents an important direction for future development. Such an approach would allow finer temporal tracking of artifact dynamics, improve real-time applicability in BCI systems, and enable integration with more advanced neural decoding. Additional research may also explore hybrid generative approaches that integrate ICA-derived components with deep generative models or multimodal EMG–EEG representations, providing even richer and more flexible models of artifact variability.

In summary, the proposed SSD framework offers a robust, scalable, and physiologically informed solution for generating multi-artifact EEG, enabling high-performance multi-label artifact classification and improving the reliability of automated preprocessing. By more accurately reproducing the complexity of real-world EEG contamination, this work contributes a significant step toward building fully automated, artifact-aware pipelines suitable for neuroscience, clinical diagnostics, and brain–computer interface applications.

\bibliography{Reference}
\end{document}